\documentclass[12pt]{article}

\usepackage{float}
\usepackage{graphics, graphicx}
\usepackage{amsfonts}
\usepackage{ natbib}

\usepackage{listings, verbatim}
\usepackage{amsmath, amssymb, amsthm, graphicx, subfigure}

\newcommand{\indicator}{\mathbf{1}}
\newcommand{\tqgamma}{\tilde q_\eta}
\newcommand{\Zset}{\mathcal{Z}}
\newcommand{\Xset}{\mathcal{X}}
\newcommand{\Rset}{\mathcal{R}}

\newcommand{\re}{\mbox{E}}

\newcommand{\beq}{\begin{equation}}
\newcommand{\eeq}{\end{equation}}
\newcommand{\beqn}{\begin{eqnarray}}
\newcommand{\eeqn}{\end{eqnarray}}

\newcommand{\cas}{\mbox{${\cal{S}}$}}

\begin{document}

\title{A Mixture-Based Approach to Regional Adaptation for MCMC}

\author{{\bf Radu V. Craiu}\\
Department of Statistics, University of Toronto\\
and\\
{\bf Antonio F. Di Narzo }\\
Department of Statistics, University of Bologna}

\maketitle

\begin{abstract}

  Recent advances in adaptive Markov chain Monte Carlo (AMCMC) include the need
  for regional adaptation in situations when the optimal transition kernel is
  different across different regions of the sample space. Motivated by these
  findings, we propose a mixture-based approach to determine the partition
  needed for regional AMCMC.  The mixture model is fitted using an online EM
  algorithm \citep[see][]{Andrieu:2006rt, capmoul1} which allows us to bypass
  simultaneously the heavy computational load and to implement the {\em regional
    adaptive algorithm with online recursion (RAPTOR)}. The method is tried on
  simulated as well as real data examples.

\end{abstract}

{\bf Keywords:} {\em Adaptive MCMC, regional adaptation, online EM, mixture model.}

\section{Introduction}

In recent years, the Markov chain Monte Carlo  (MCMC) class of computational algorithms has been enriched with adaptive MCMC (AMCMC). Spurred by the seminal paper of \cite{Haario:1999kx} an increasing body of literature has been devoted to the study of AMCMC.  It has long been known that the fine tuning of the proposal distribution's parameters in a Metropolis sampler is central to the performance of the algorithm.  \cite{Haario:1999kx}, \cite{Haario:2005fj}, \cite{androb},  \cite{Andrieu:2006rt}, \cite{Andrieu:2005lq} and \cite{MR2340211} have  provided the theory needed to prove that it is possible to adapt the parameters of the proposal distribution "on the fly", i.e. while running the Markov chain and using for tuning the very samples produced by the chain. The AMCMC algorithms may be vulnerable to the  multimodality of the target distribution and more care needs to be taken in implementing the AMCMC paradigm. 
In \cite{craiu-jeff-yang} a few possible approaches are discussed, central among which is the regional adaptive algorithm (RAPT) designed for  Metropolis samplers. However, the premise for RAPT  is that a partition of the sample space is 
given and it is approximately correctly specified.  While sophisticated methods exist to detect the modes of a multimodal distribution  \citep[see][]{smin1,smin2,MR1837132} it is not obvious how to use such techniques for defining the desired partition of the sample space. We follow here the 
methods of  \cite{Andrieu:2006rt} and
\cite{capmoul1} to propose a mixture-based approach for adaptively determining the 
boundary between high probability regions.  We approximate the target distribution using a mixture of Gaussians whose parameters are used to define the partition. The theoretical challenges lie in the fact that the volume of data used for fitting the mixture increases as the simulation progresses and the data is not independent since it is made of realizations of a Markov chain. Both 
challenges have been tackled   by \cite{Andrieu:2006rt} and
\cite{capmoul1}.

In the next section we briefly review the RAPT algorithm and the online EM
algorithm of \cite{capmoul1}.  In section 3 we describe the methodology behind
the regional adaptive algorithm with online recursion (RAPTOR). The simulation
studies and real data application are discussed in Sections 4 and 5,
respectively.

\section{Regional Adaptation and Online EM}

\subsection{Regional Adaptation (RAPT)}

Regional adaptation is motivated by the  fundamental and natural idea that, in many situations,  the optimal 
proposal distribution used in a Metropolis sampling algorithm may be different in separate regions of the sample space  $\cas$.   For now, assume that we are {\em given} a  partition of the space
$\cas$ made of  two regions $\cas_{1},\cas_{2}$. The mixed RAPT algorithm 
for a random walk Metropolis (RWM) sampler uses the following mixture as a proposal distribution
\beq
\label{mrapt}
 Q(x,dy)=(1-\beta) \sum_{i=1}^2 1_{\cas_i}(x)[\lambda_1^{(i)}Q_1(x,dy)+\lambda_2^{(i)}Q_2(x,dy)] + \beta Q_{whole}(x,dy),
 \eeq
where $Q_i$ is adapted using samples from $\cas_i$ and $Q_{whole}$ is adapted using all the samples in $\cas$.
The mixing parameters $\lambda_1^{(i)}$, $i=1,2$ are also adapted while the parameter   $\beta$ is constant throughout the simulation. Details regarding the adaptation procedures for the above distributions and parameters can be found in  \cite{craiu-jeff-yang} who also provide a proof regarding the asymptotic convergence of the algorithm.

One can see that, regardless of the region the chain is currently in, the proposal distribution is a mixture of three distributions: $Q_1, Q_2$ which are approximately optimal choices for the target restricted to $\cas_1$ and $\cas_2$, respectively, and $Q_{whole}$, which has the purpose of ensuring good traffic between the two regions. The reason we use a mixture with these three components (as opposed to using a mixture with the components $Q_i$ and $Q_{whole}$ when the chain is in $\cas_i$)  is intuitively motivated by the uncertainty of 
determining  the ideal partition $\cas=\cas_1 \cup \cas_2$.  The degree of success for RAPT depends on whether the partition used is a relatively good approximation of the ideal one. In the next section we propose to adaptively modify the partition between the two regions using the online EM algorithm.

\subsection{Online EM}\label{sec:online-em}

Denote $\pi$  the target distribution of interest.  Working under the assumption that $\pi$ is multi-modal one 
can try to approximate $\pi$ using a mixture of Gaussian distributions. The approximation is in many cases accurate once the distribution $\pi$ can be well approximated by a Gaussian in a neighborhood of each local mode.  The analysis of mixture models has relied for a while now on the EM algorithm  \citep{dlr} as discussed by  \cite{titt} and references therein.  In the MCMC setup the amount of data available to fit the mixture increases as the simulation progresses, therefore making unfeasible the traditional implementation of the algorithm. An added complication is that  the streams of data  contain dependent realizations as they are produced by running one or more Markov chains.    Both difficulties are dealt with effectively by  \cite{Andrieu:2006rt}  who propose an online EM algorithm that updates the parameter estimates as more data become available. The algorithm is further refined by  \cite{capmoul1}.

The  M-step for the classical EM algorithm involves the maximization (in $\theta$)
of 
$$ Q_{\theta'}(y_{1:n};\theta)=\sum_{i=1}^n E[\log f(X_i;\theta)|\theta',y_i]$$
where  $Y_{1:n}$ are the $n$-dimensional observed data and $X_i$ is the $i$-th unit  complete data.

The online EM  of \cite{Andrieu:2006rt}  modify the $Q$ function to
\beq
\label{q-onlem}
\hat{Q}_{n+1}(\theta)= \hat{Q}_{n}(\theta)+\gamma_{n+1} \left (\re_{\hat{\theta}_n}[ \log f(X_{n+1};\theta)|Y_{n+1}] -\hat{Q}_{n}(\theta) \right )\eeq
and set $\hat{\theta}_{n+1}$ as its maximizer. Here $n$ is the size of the sample $y_{1:n}$  available at the $n$-th iteration. Note that the volume of available data increases at  each iteration of the algorithm while the weights $\gamma_n$ are set to decrease with $n$.  For additional details we refer the reader to 
\cite{Andrieu:2006rt} and \cite{capmoul1}.

\section{Mixture based boundary adaptation}

\subsection{An illustrative example}
\begin{figure}
  \centering \subfigure[Target prob. density]{
    \includegraphics[width=0.3\textwidth]{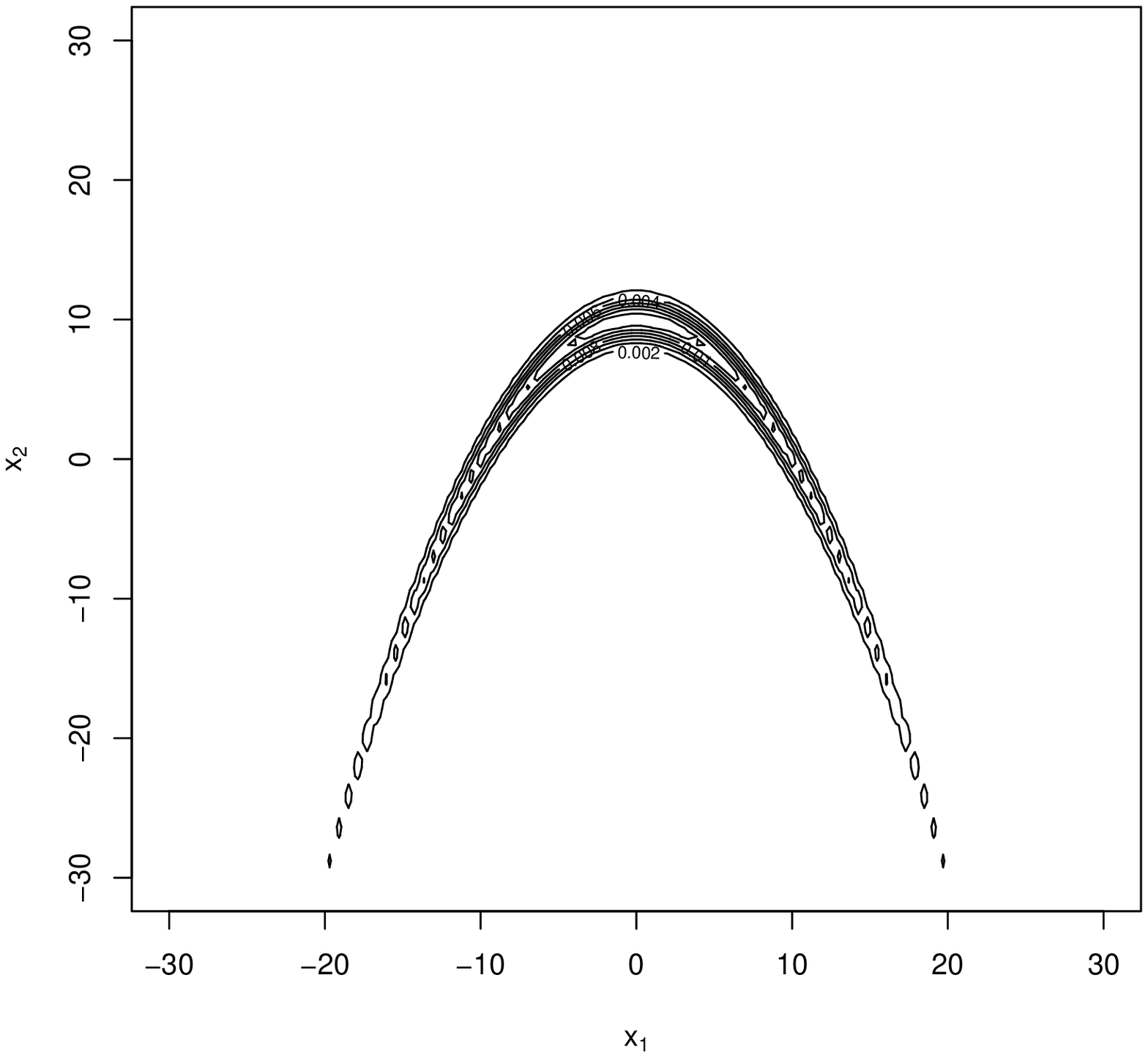}
    \label{fig:bananaEx1Contour}}
  \subfigure[Chains scatterplot]{
    \includegraphics[width=0.3\textwidth]{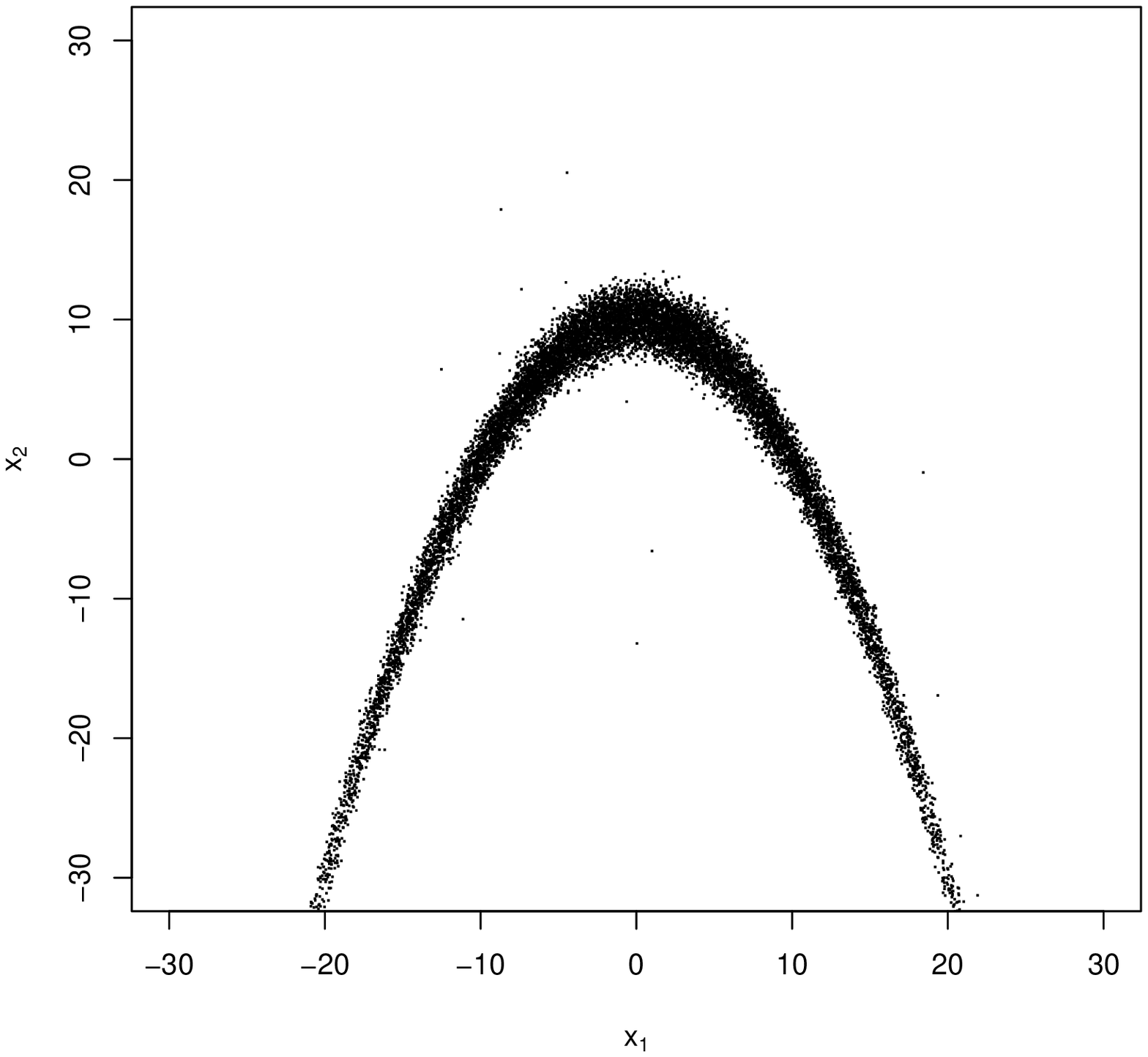}
    \label{fig:bananaEx1Cloud}}
  \subfigure[Final mixture estimate]{
    \includegraphics[width=0.3\textwidth]{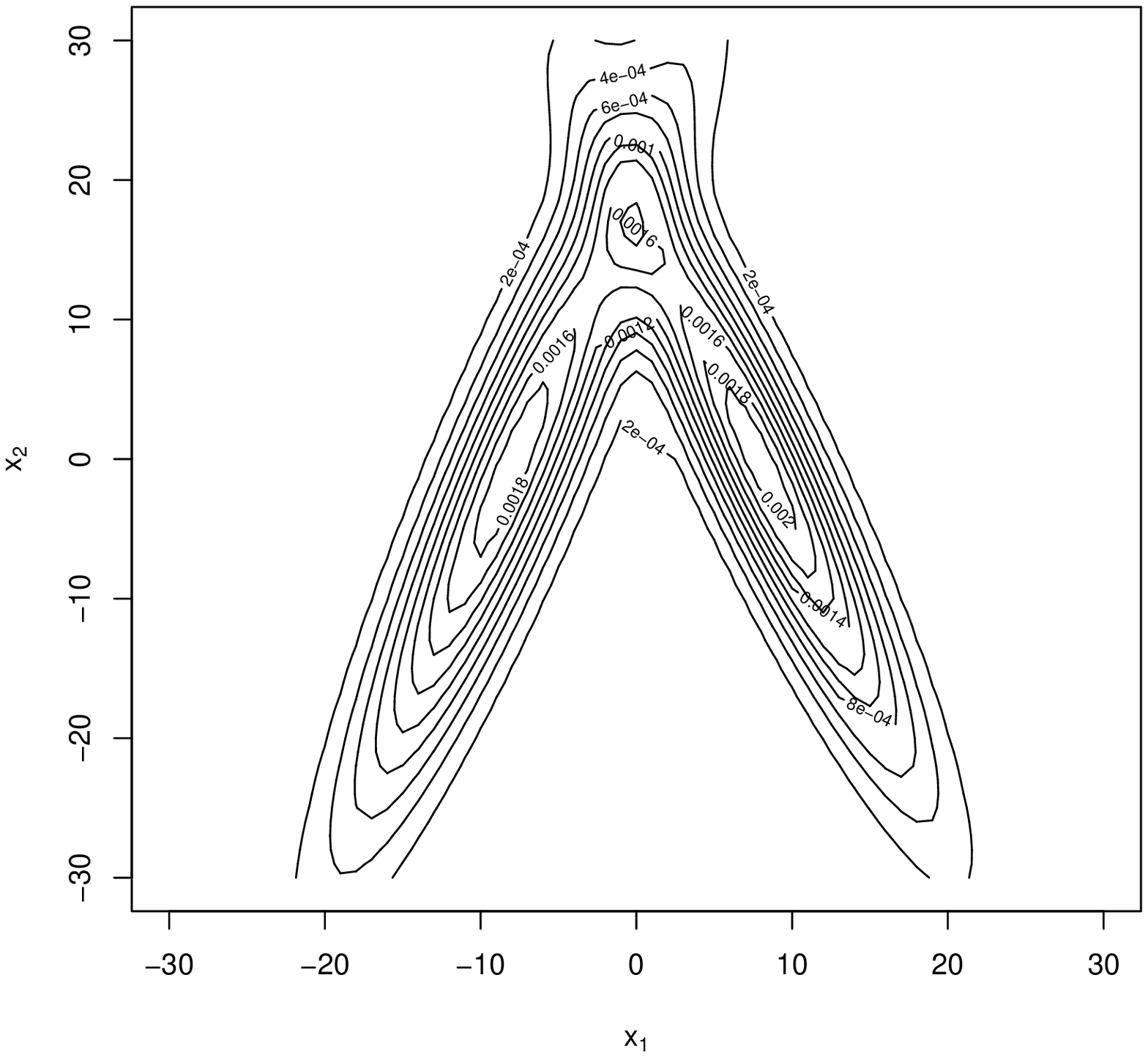}
    \label{fig:bananaEx1MixFit}}
  \caption{{\it Example of regional adaptation applied to a curved target
    distribution.}}
  \label{fig:bananaRaptorOverview}
\end{figure}

Consider the curved density of general form \citep[see][]{adaptex}:
\begin{equation}\label{eq:bananaDensity}
  f(x; B) \propto \text{exp} \left [ - x_1^2/200 - \frac{1}{2} (x_2 + B x_1^2 - 100 B)^2
  - \frac{1}{2}(x_3^2 + x_4^2 + \ldots + x_d^2) \right ]
\end{equation}
For illustration, we consider here the $2$-dimensional version
of~\eqref{eq:bananaDensity} and in section~\ref{sec:cur-tar-distr} we will
perform a simulation study for the $5$-dimensional version
of~\eqref{eq:bananaDensity}.  In Figure~\ref{fig:bananaEx1Contour} the contour
plot for $B=0.1$ is shown.  The correlation between the two coordinates is close
to $0$, so a standard adaptive RWM algorithm may use a nearly
spherical, largely overdispersed Gaussian distribution.

Within the RWM framework we can gain efficiency by splitting
the state space horizontally into two regions, and adapting the covariance
matrices in each region. Using the online EM algorithm for the MCMC sampling output, we can  fit a mixture of distributions to
$\pi$, and adapt the chain's transition kernel
according to the mixture parameter values.

We run $10$ parallel chains of our RAPTOR algorithm for $25000$ iterations,
using the first $1000$ as burn-in and allowing exchange of information between
chains \citep[see][]{craiu-jeff-yang}.  In Figure~\ref{fig:bananaEx1Cloud} we show the
scatterplot of the  values  obtained using all the ten chains.  The final Gaussian
mixture estimate is plotted in Figure~\ref{fig:bananaEx1MixFit}. Here we can see
that the final mixture fit mimics well the target density. 

We design RAPTOR so that it exploits the Gaussian mixture approximation and 
increases the sampling efficiency while adding  little computational overhead. In the next section 
we discuss  how the Gaussian mixture approximation can be used to: i)  define a convenient
partitioning of the state space and ii)  tune the proposal distribution of  the
Metropolis sampler within each region.

\subsection{RAPT with online recursion}

Consider the $K$ components mixture model:
\begin{equation}
  \label{eq:mixtureModel1}
  \tqgamma(x) = \sum_{k=1}^K \beta_\eta^k N(x; \mu_\eta^k, \Sigma_\eta^k)
\end{equation}
where $N(; \mu, \Sigma)$ is the probability density of a Gaussian distribution
with mean $\mu$ and covariance matrix $\Sigma$.  In standard mixture modelling
terminology, \eqref{eq:mixtureModel1} is called the `incomplete' likelihood, and
the complete likelihood is written as follows:
\begin{equation}
  \label{eq:mixtureModelComplete}
  f_\eta(x, z) = \prod_{k=1}^K \left [\beta_\eta^k N(x; \mu_\eta^k, \Sigma_\eta^k)\right ]^{\mathbf{1}(z=k)}
\end{equation}
where $z$ is an unobserved labelling variable taking values in the finite set
$\{1, 2, \ldots, K\}$.  For notational convenience all the parameters involved
in the model are included in the vector $ \eta = \{(\beta_\eta^k, \mu_\eta^k, \Sigma_\eta^k),
k=1, \ldots, K\}$, with $\eta \in \Omega$.  We propose to approximate the target
distribution $\pi$ with $\tqgamma$ so that the Kullback-Leibler distance between
$\pi$ and $\tqgamma$ is minimized.  Given the
approximation~\eqref{eq:mixtureModel1} to $\pi$ we define the region
$\cas_\eta^k$ as the set in which the $k$-th component of the mixture density
$\tqgamma$ dominates the other ones., i.e.
\begin{equation}
  \label{eq:boundary1}
  \begin{aligned}
    \cas_\eta^k = \{x: \; \text{arg max}_{k'} N(x; \mu_\eta^{k'},
    \Sigma_\eta^{k'}) = k\}.
  \end{aligned}
\end{equation}
The implicit assumption is that, in $S_\eta^k$,  $\pi$ is well approximated by a Gaussian distribution
with mean $\mu_\eta^k$ and  covariance matrix
$\Sigma_\eta^k$.  This approximation can be exploited in the definition of the
local Metropolis proposal distribution.

Note that the mixture parameters $\beta_\eta^k$ are omitted from the boundary
definition~\eqref{eq:boundary1}. We also do not exclude components with small weights. It should be also noted that in the current approach $K$ is fixed and its choice can be based on an exploratory numerical analysis of the target $\pi$ (e.g., the number of local maxima of $\pi$).

We expect that  the  recurrent update of the boundary between regions will 
eventually lead to  regions that are optimal or close to optimal. Perhaps more importantly, this approach provides a general strategy to tackle the tricky issue of partitioning the sample space.  Although in principle we could continue to use the proposal distribution  (\ref{mrapt}), a good partition of the sample space allows the use of 
\begin{equation}
  \label{eq:raptorKernel1}
  Q_\eta(x, dy) = ( 1 - \alpha ) 
  \sum_{k=1}^K 1_{\cas_\eta^k}(x) N(y; x, \epsilon_d \Sigma_\eta^k )dy,
  + \alpha N(y; x, \epsilon_d \Sigma_\eta^w )dy
\end{equation}
where $\Sigma_\eta^w$ is the marginal variance of $\tqgamma$, $\epsilon_d =
2.38^2/d$, a choice based on the optimality results 
obtained for the RWM  by~\cite{rgg} and \cite{MR1888450}, and $\alpha\in(0,1)$ is a fixed
weight which controls the flow between regions.

The transition kernel~\eqref{eq:raptorKernel1} depends on the mixture parameters $\eta$
in two ways: via the regions definition \eqref{eq:boundary1} and, more directly,
via the covariance matrices $\Sigma_\eta^k$ and $\Sigma_\eta^w$. The
adaptation strategy consists in replacing at each iteration, say $n$th, the parameter
$\eta$ with an estimate $\eta_n$ which is obtained from the chain's realizations observed so
far.

\begin{figure}
  \centering
  \subfigure[Different means, equal variances]{\label{fig:RAPTORRegionsShapesA}
    \includegraphics[width=0.3\textwidth]{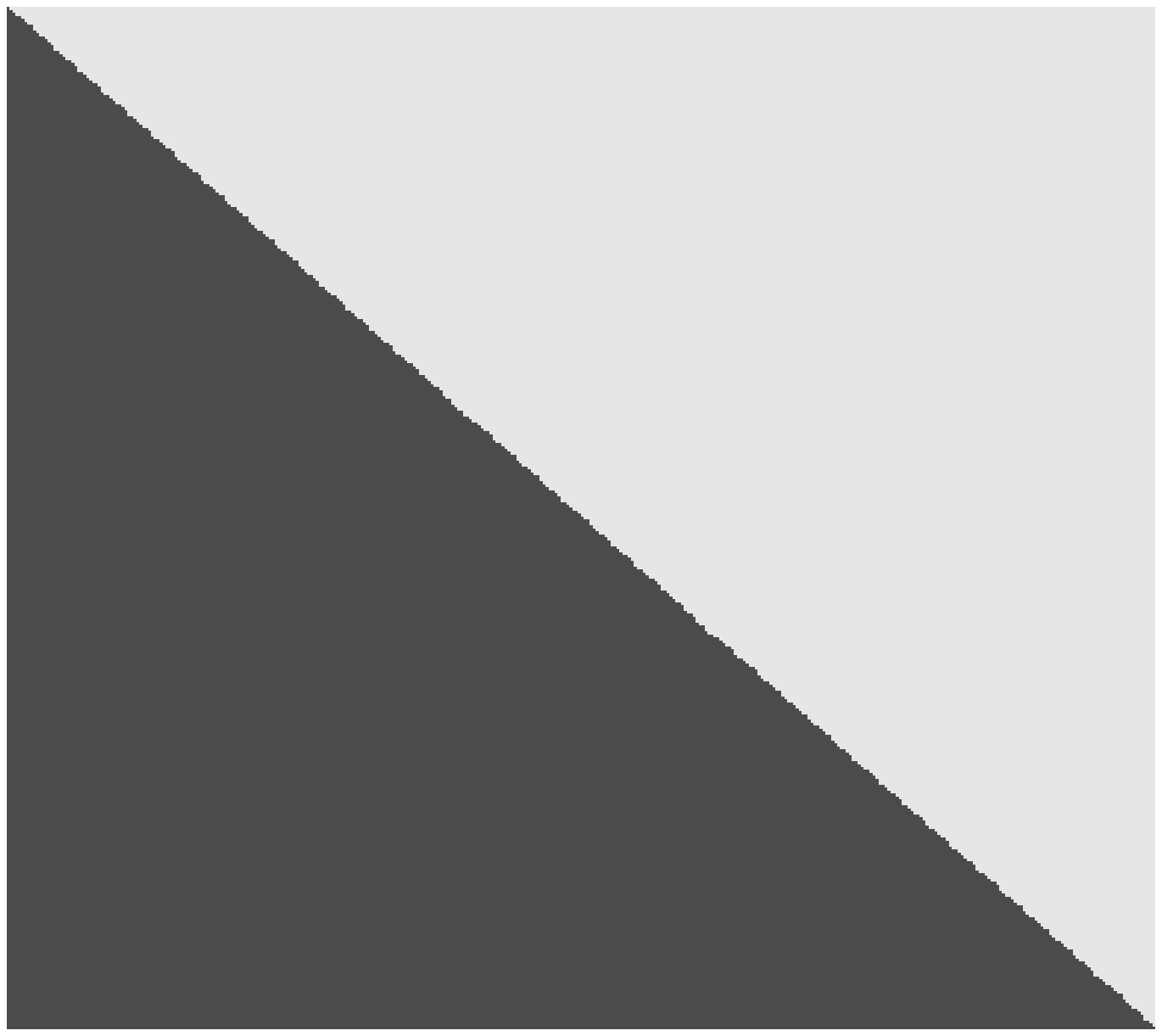}}
  \subfigure[Different means, different variances]{\label{fig:RAPTORRegionsShapesB}
    \includegraphics[width=0.3\textwidth]{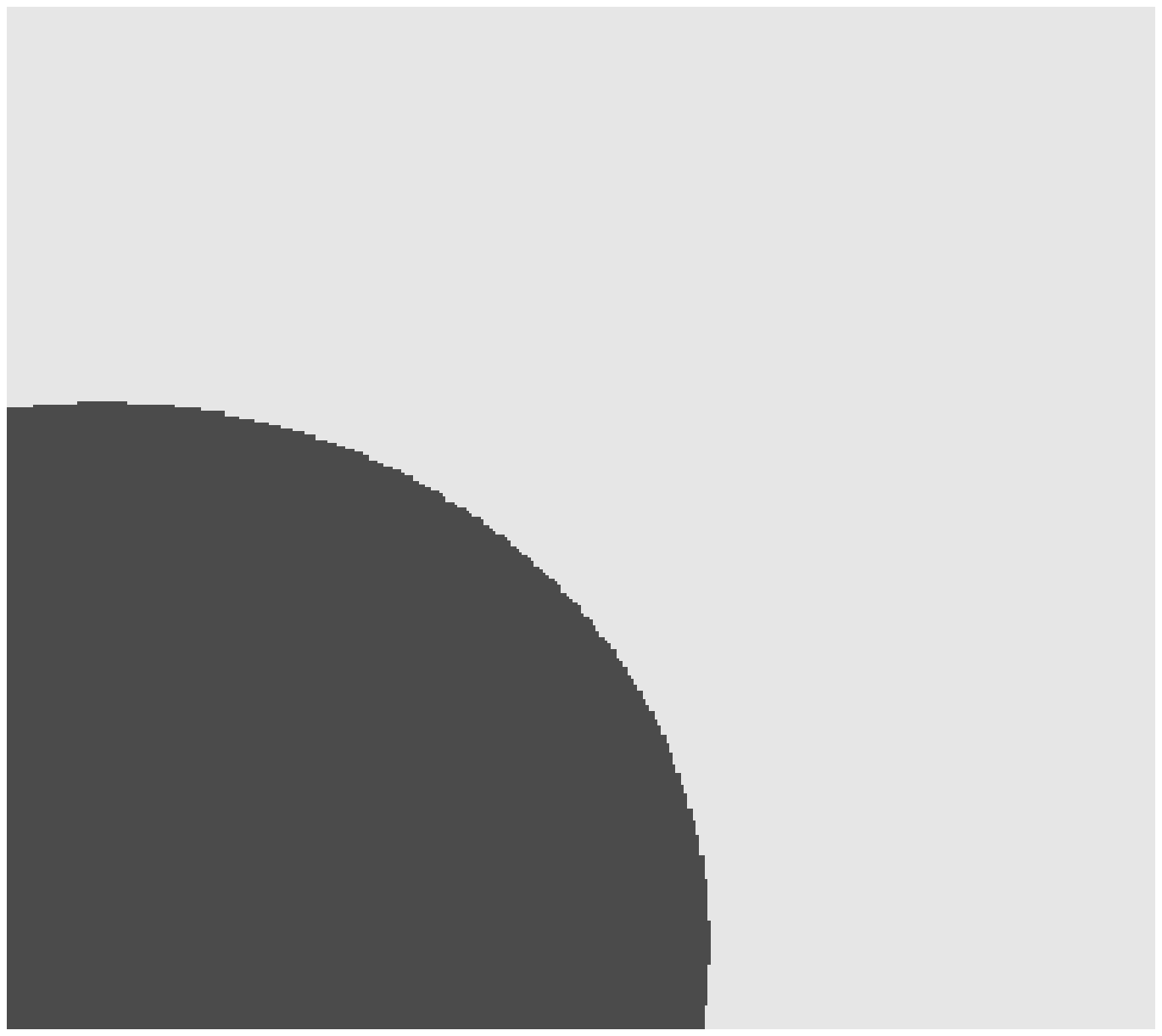}}
  \subfigure[Equal means, different variances]{\label{fig:RAPTORRegionsShapesC}
    \includegraphics[width=0.3\textwidth]{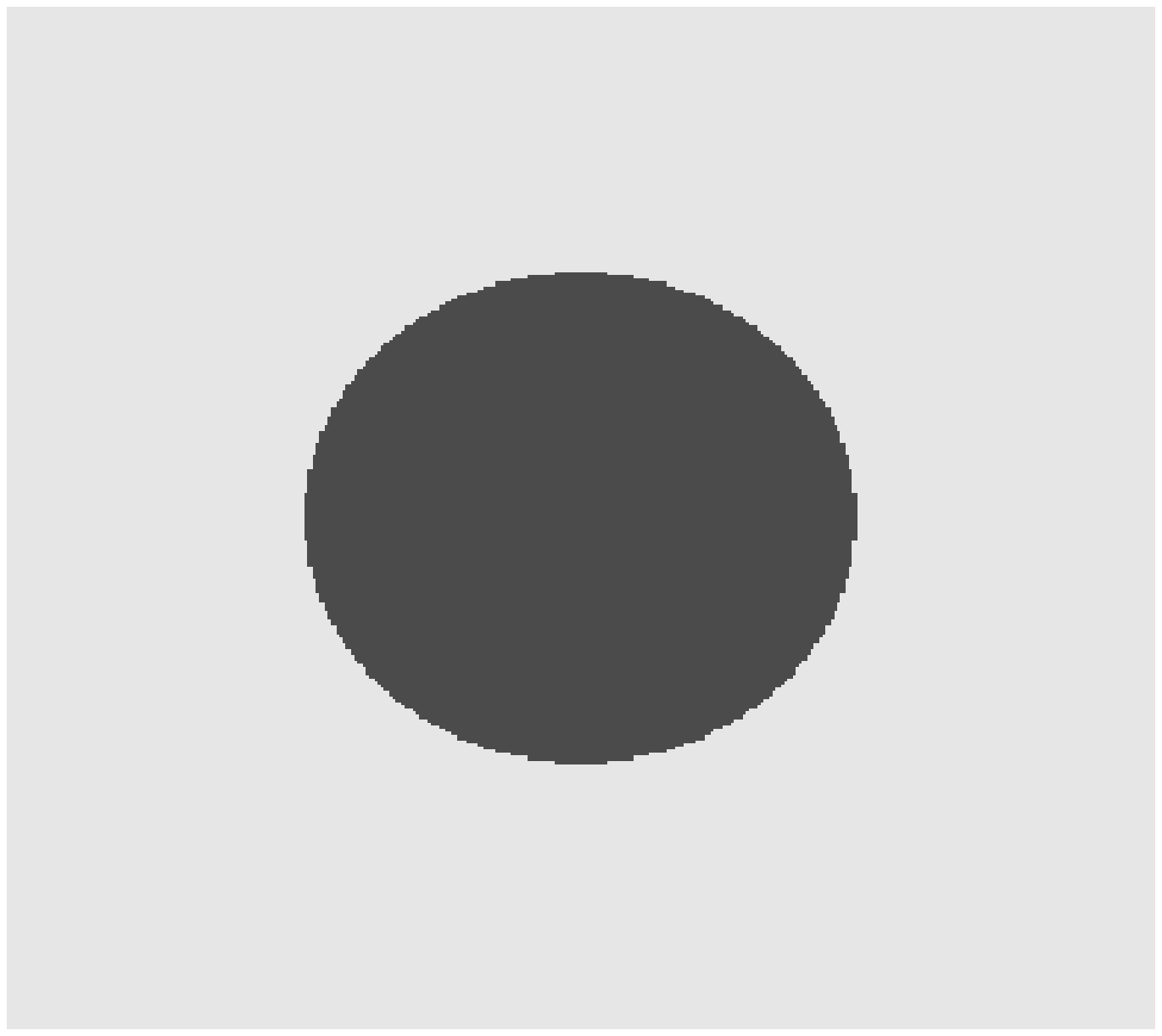}}
  \caption{{\em  RAPTOR-defined regions  for different relative values of the mixture
    components parameters. Region $1$ in dark gray, region $2$ in light gray.}}
  \label{fig:RAPTORRegionsShapes}
\end{figure}

In Figure~\ref{fig:RAPTORRegionsShapes} some actual shapes of the boundary
between two regions as specified by~\eqref{eq:boundary1} are shown. It can be
seen that the boundary has a good level of flexibility, and can represent both
convex and concave regions. Indeed, regions can also have `holes'
as seen in Figure~\ref{fig:RAPTORRegionsShapesC}.

\subsection{The online EM for RAPTOR}
\label{sec:rec-strat-mixt-par-est}
If $\nu^k_i=P(Z_i=k|x_i,\eta_{i})$ then
\begin{equation}
  \label{eq:w_k_i}
  \nu^k_i = \frac{\beta_{i-1}^k \phi(x_i; \mu_{i-1}^k, \Sigma_{i-1}^k)}
  {\sum_{k'} \beta_{i-1}^{k^\prime} \phi(x_i; \mu_{i-1}^{k^\prime}, \Sigma_{i-1}^{k^\prime})},
\end{equation}
where $\eta_n = \{(\beta_n^k, \mu_n^k, \Sigma_n^k), k
= 1, \ldots, K\}$.  If we define $s_n^k$
\begin{equation}
  \label{eq:s_k_n}
     s_n^k  = \frac{1}{n} \textstyle\sum_{i=1}^n \nu_i^k  = (1 - 1/n) s_{n-1}^k + 1/n\; \nu_n^k,
\end{equation}
then the  recursive estimator
$\eta_n = \{ (\beta_n^k, \mu_n^k, \Sigma_n^k): k=1,\ldots,K \}$
is
\begin{equation}
  \label{eq:gamma_n}
  \begin{aligned}
    \beta_n^k & = s_n^k, \\
    \mu_n^k & = \frac{1/n \; \sum_{i=1}^n \nu_i^k x_i}{s_n^k}, \\
    \Sigma_n^k & = \frac{1/n\; \sum_{i=1}^n \nu_i^k x_i x_i'}{ s_n^k }
    - \mu_n^k \mu_n^{k\prime}.
  \end{aligned}
\end{equation}

The scheme \eqref{eq:w_k_i}-\eqref{eq:gamma_n} defines an  online EM whose convergence 
has been proved by~\cite{Andrieu:2006rt}. They have shown that, under mild regularity conditions on $\pi$, the
estimator defined by \eqref{eq:w_k_i}-\eqref{eq:gamma_n} converges to the value
of $\eta$ which minimizes the Kullback-Leibler divergence between $\pi$ and
$\tqgamma$.   Moreover,~\cite{Andrieu:2006rt} proved also the ergodicity of an adaptive  independent Metropolis sampler whose proposal parameters are the estimates produced by the online EM.  
The detailed derivation of  equations \eqref{eq:w_k_i}-\eqref{eq:gamma_n} is shown in  appendix ~\ref{sec:appendix:-line-em}.

We should note that Remark 8 in \cite{Andrieu:2006rt} points out the direct extendability of their proof  to the current  RWM setting. Therefore, we do not replicate the proofs here and refer the reader to \cite{Andrieu:2006rt} for the theoretical groundwork.

\subsubsection{Inter-Chain Adaptation extension}\label{sec:INCAExtension}
The recursive estimation scheme defined above can be easily extended to
the context of multiple parallel chains, allowing inter-chain adaptation \citep[INCA, see][]{craiu-jeff-yang}.

Denote the $MN$ samples obtained from $M$ parallel chains by $\{\{X_t^m\}, 1\le
m \le M, \ 1\le t \le N\}$. For each $N$, we can build a pooled chain $\{Y_k\}$
using, for any $1\le k \le MN$, $Y_k=X_{j(k)}^{i(k)}$, where $i(k)=k-M [j(k)-1]$
and $j(k)=\lfloor {k+M-1 \over M} \rfloor$.  We apply the recursive estimation
scheme \eqref{eq:gamma_n} to the sequence $\{Y_k\}_k$ without modifications.

\section{Simulations}
In this section, we illustrate the performance of the  RAPTOR algorithm using  Gaussian mixtures  under different scenarios designed to cover 
a wide range of possibilities. In addition, we test RAPTOR on an irregularly shaped target distribution which has been
already studied in \cite{Haario:1999kx} and \cite{adaptex}. In this scenario,
the target probability density has only one mode and the domain is well
connected, so that there is no real risk for a standard Metropolis algorithm of
remaining trapped in one region of the state space. However, we will show that
even in such cases regional adaptation, in particular RAPTOR, improves over the non-regional Adaptive Metropolis algorithm.

\subsection{Algorithms comparison}
\label{sec:algor-comp}
In the following, we will compare different Metropolis algorithms using the following summaries:
\begin{itemize}
\item[(I)] Acceptance Rate (AR),
\item[(II)] Mean Squared Error (MSE) of the sample mean estimator,
\item[(III)] Bias of the sample mean estimator,
\item[(IV)] Distance between the target cumulative distribution function (CDF) and the empirical cumulative distribution function (ECDF) .
\end{itemize}

We propose to use (IV) as a more comprehensive indicator of the sampling efficiency, compared to (III) which  summarizes only the first two moments of the Monte Carlo estimator. Evidently, the main caveat of (IV) is that it cannot be used in real applications when the target CDF is not known.

For numerically evaluating the distance between an ECFD calculated using the MCMC output  \citep[see][]{sensing, chen-sha-bra}
and 
 the  target CDF, we introduce the index
\begin{equation}
  \label{eq:Dn1}
  D_n = \int |F_n - F|^2 dF,
\end{equation}
where $F_n$ is the ECDF obtained using
$\{X_t\}_{1\le t\le n}$, i.e.,
\begin{equation}
  \label{eq:Fn}
  F_n(z) = \frac{1}{n} \sum_{t=1}^n \indicator\{X_t \leq z\}.
\end{equation}
In cases where it's easy to get i.i.d. samples from $F$, the integral in
\eqref{eq:Dn1} can be computed numerically by Monte-Carlo simulation. More precisely, 
given a set $\{y_1, \ldots, y_M\}$ of i.i.d. draws from $F$, we approximate $D_n$
using
\[ \hat D_n = \frac{1}{M} \sum_{j=1}^M |F_n(y_j) - F(y_j)|^2 . \] Note that, in
the above formula, the algorithm under evaluation is involved through the ECDF
$F_n$, while the target CDF is used both in $F$ and in the generation of the
sample $\{y_j\}_j$. The encompassing nature of the index is obvious, as in practice the objective of the MCMC procedure is precisely to get
good samples from $F$.  Moreover, the measure $D_n$ is appealing  because 
by integrating with respect to  $F$ we give more weight to regions of the
state space with higher probability, and automatically ignore discrepancies
between $F_n$ and $F$ in zones which are of low interest.

For simplicity, we will use the notation $D_n$ even when
its Monte-Carlo approximation is used instead. In practice, we will report  $\bar{D}_n$, the 
average of $B$ independent replicates of $D_n$, i.e.
\begin{equation}
  \label{eq:Dn2}
  \bar D_n = \frac{1}{B} \sum_{b=1}^B D_n^{(b)}.
\end{equation}

\subsection{Gaussian mixture target distribution}
In this section the  target distribution is a Gaussian mixture 
\begin{equation}
  \label{eq:1}
  f(x; \xi, d, S) = \xi N(x; -d \times \mathbf{1}, \text{I}_5) + (1-\xi) N(x; d \times
  \mathbf{1}, S \times \text{I}_5),
\end{equation}
where $\xi, d, S \in \mathbf{R}$ and $N(; \mu, \Sigma)$ is the probability density
of a $5$-dimensional Gaussian distribution with mean $\mu$ and covariance matrix
$\Sigma$.   For increasing
values of $d$, the target distribution presents two modes which are more and
more separated and $S$ is the ratio between the marginal variances of the two  mixture components. A priori, we expect RAPTOR to make a difference when $d$ is at least moderately large.

We compare 4 different adaptive RWM algorithms:
\begin{itemize}
\item RAPTOR
\item RAPT, with boundary $\{x_1 + x_2 = 0\}$
\item RAPT, with boundary $\{x_1 + x_2 = 2\}$ (named RAPT2 in the following)
\item Adaptive Metropolis (AM)~\citep{Haario:1999kx}
\end{itemize}

We have run $10$ chains in parallel with randomized starting values, each for a
total of $10000$ iterations, using the first $5000$ as a burn-in, and allowing
sharing of information between chains (see sec.~\ref{sec:INCAExtension}). The
simulation has been replicated $200$ times. Initial values for local means and
covariance matrices have been set as follows:
\begin{equation}
  \label{eq:gaussMixStartingValues}
    \mu_0^k  = 1.5 \times \mu^k_\text{true}, \
    \beta_0^k  = 0.5, \
    \Sigma_0^k  = 0.5 \times \Sigma^k_\text{true}, \
    \Sigma_0^w  = 5.0 \times \Sigma^2_\text{true}
\end{equation}
For all algorithms, these values have been used for setting the starting
proposal covariance matrices. For RAPTOR, these have been also used as starting
parameters estimates.  In each simulation and for each algorithm we report the
mean squared error (MSE) and the acceptance rates. These were computed based on
the $200$ replications of the simulation. The results are reported in Table
\ref{tab:gaussMixSimul}.

\begin{table}
  \centering
  \begin{tabular}{lrrcrr}
    \hline \hline
    & \multicolumn{2}{c}{\textbf{d=3, S=1}} &\multicolumn{1}{c}{}& \multicolumn{2}{c}{\textbf{d=0,
        S=4}}
    \tabularnewline \cline{2-3} \cline{5-6}
    \textbf{Algorithm} & \multicolumn{1}{c}{\textbf{AR}} &
    \multicolumn{1}{c}{\textbf{MSE}} &
    &
    \multicolumn{1}{c}{\textbf{AR}} &
    \multicolumn{1}{c}{\textbf{MSE} $\times 100$} \\
    \hline
    RAPTOR & 0.2485 & 0.0813 && 0.3092 & 0.1888\\
    RAPT & 0.2477 & 0.1239 && 0.2747 & 0.2410\\
    RAPT2 & 0.2430 & 0.1309 && 0.2687 & 0.3346 \\
    AM & 0.0937 & 0.1671 && 0.2739 & 0.5837\\
    \hline
  \end{tabular}
  \caption{{\em Gaussian mixture target distribution: MSE and acceptance rates in two
  different scenarios, $\xi = 0.5$.}}
  \label{tab:gaussMixSimul}
\end{table}

For $\xi=0.5$, $d=3$ and $S=1$, all the regional adaptive algorithms reach an
average acceptance rate of around $24\%$ while AM remains below $10\%$. Also in
terms of MSE, all the regional adaptive algorithms outperform the simple
adaptive Metropolis. However, here we see that RAPT with the boundary $\{x_1
+ x_2 = 0\}$ has a slightly smaller MSE than RAPT2 which uses the
boundary $\{x_1 + x_2 = 2\}$, and that RAPTOR performs better than both,
lowering again MSE by more than $30\%$.

Encouraging results have been obtained also for the scenario with two identical
target mixture means, same weights, but different variances. Here the optimal
mixture-based boundary has the shape showed in
Figure~\ref{fig:RAPTORRegionsShapesC}, so that local RAPTOR proposals have
smaller steps in the center of the distribution and bigger steps in the
tails. The global proposal induces jumps with a length  between the lengths  produced by the
two local proposals.  In this scenario, the boundary produced using  RAPTOR  
differs drammatically from that of RAPT and RAPT2, and this yields an efficiency gain resulting in 
a $21\%$ decrease in the MSE of the sample mean estimator and a $12.5\%$
improvement of the average acceptance rate over RAPT.

\subsection{A curved target distribution}\label{sec:cur-tar-distr}
We consider the probability density given in equation~\eqref{eq:bananaDensity} in the case of $5$ dimensions.
We run each chain $500$ times, with starting conditions randomly drawn from a
uniform distribution on the hypercube $(-2; 2)^5$. For all methods, we used the
first $4000$ iterations as a burn-in.  We compare again the same four different
algorithms:
\begin{itemize}
\item RAPTOR
\item RAPT, with boundary fixed to $\{x_1=0\}$
\item RAPT, with boundary fixed to $\{x_2=-1\}$ (named RAPT2 in the
  following)
\item AM
\end{itemize}

For all the algorithms, starting parameters values were determined on the basis
of a preliminary simulation stage, common to the $500$ replications. We have run
$4000$ iterations of a Gaussian Metropolis Random Walk to get initial estimates
of the target distribution covariance matrix, as well as initial estimates for a
Gaussian mixture approximation, obtained by running a classical EM algorithm.
The weight $\alpha$ have been set to $0.2$ in RAPTOR as well as in both RAPT
implementations.

\begin{table}
  \centering
  \begin{tabular}{rrrr}
    \hline \hline
    RAPTOR & RAPT & RAPT2 & AM \\
    \hline
    11.23 &  11.22 & 8.81 & 4.84 \\
    \hline
  \end{tabular}
  \caption{{\em Curved target distribution simulation: acceptance rates (\%) averaged
  over 500 independent runs of each chain}}
  \label{tab:bananaDim5AR}
\end{table}

\begin{table}
  \begin{center}
    \begin{tabular}{lrrr}\hline\hline
      &
      \multicolumn{1}{c}{$\bar X_1$}&
      \multicolumn{1}{c}{$\bar X_2$}&
      \multicolumn{1}{c}{$\bar X_3$}\tabularnewline
      \hline
      RAPTOR&$4.1229$&$ 8.7565$&$0.0099$\tabularnewline
      RAPT&$4.6342$&$ 8.2658$&$0.0098$\tabularnewline
      RAPT2&$4.0647$&$17.3305$&$0.0146$\tabularnewline
      AM&$4.1506$&$11.7487$&$0.0235$\tabularnewline
      \hline
    \end{tabular}
  \end{center}
  \caption{{\em Curved target distribution simulation: MSE of the estimator of the
    mean of the first 3 coordinates, for each algorithm.
    Estimates are based on 500 independent chains replications.}}
  \label{tab:bananaDim5MSE_Bias}
\end{table}

\begin{figure}
  \centering
  \includegraphics[width=0.8\textwidth]{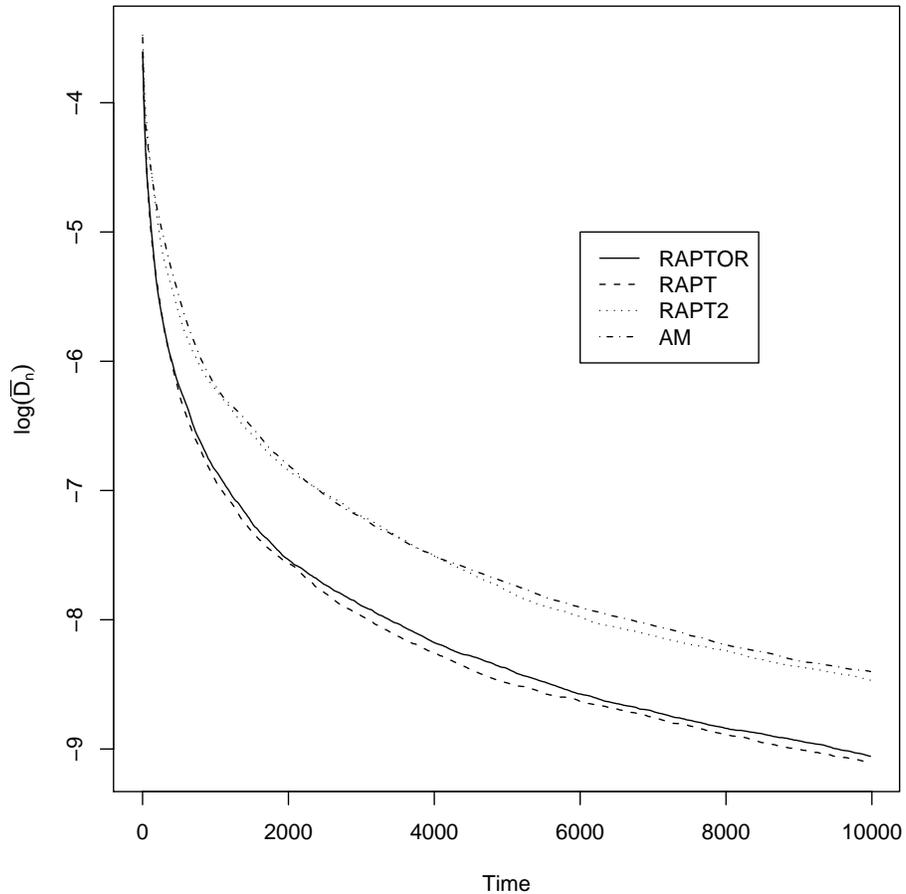}
  \caption{{\em Log-average distance from the curved target distribution}}
  \label{fig:Dn_Banana_5}
\end{figure}

\begin{figure}
  \centering
  \includegraphics[width=0.8\textwidth]{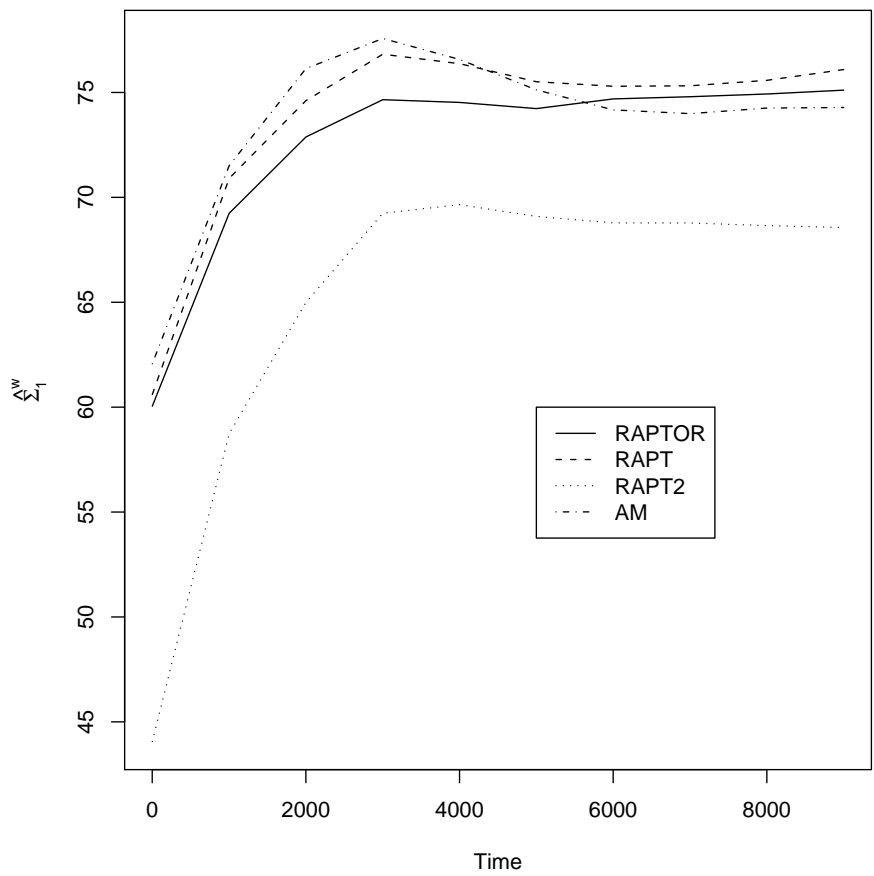}
  \caption{{\em Curved target distribution: average marginal variance estimates}}
  \label{fig:bananaMarginalVariancesEstimates}
\end{figure}

\begin{table}
  \centering
  \begin{tabular}{lrrrcrrr}\hline\hline
    \multicolumn{1}{l}{\bfseries }&
    \multicolumn{3}{c}{\bfseries Region 1}&
    \multicolumn{1}{c}{\bfseries }&
    \multicolumn{3}{c}{\bfseries Region 2}
    \tabularnewline \cline{2-4} \cline{6-8}
    \multicolumn{1}{l}{\bfseries Algorithm}&
    \multicolumn{1}{c}{$\hat\mu^1_1$}&
    \multicolumn{1}{c}{$\hat\Sigma^1_{1,1}$}&
    \multicolumn{1}{c}{$\hat\rho^1_{1,2}$}&
    \multicolumn{1}{c}{}&
    \multicolumn{1}{c}{$\hat\mu^2_1$}&
    \multicolumn{1}{c}{$\hat\Sigma^2_{1,1}$}&
    \multicolumn{1}{c}{$\hat\rho^2_{1,2}$}
    \tabularnewline \hline
    RAPTOR&$-7.103$&$25.421$&$0.944$&&$7.619$&$24.593$&$-0.952$\tabularnewline
    RAPT&$-7.159$&$24.834$&$0.951$&&$7.293$&$25.304$&$-0.951$\tabularnewline
    \hline
  \end{tabular}

  \caption{{\em Curved target distribution: average local estimates}
  \label{tab:bananaLocalEstimates}}
\end{table}

In table~\ref{tab:bananaDim5AR} we report the acceptance rates for the four
different sampling strategies. We see that all the three regional adaptive
methods outperform the simple Adaptive Metropolis. The best performance is
achieved by RAPTOR and the RAPT with the vertical boundary $\{x_1=0\}$. Indeed,
we will see that RAPTOR tends to approximate the RAPT boundary very well.

The MSE for the mean estimates shows that all the tested algorithms are roughly
equivalent in the first coordinate, while RAPTOR and RAPT achieve the best
performances on the second coordinate, with RAPT having the best score, and
RAPTOR closely following. The results also emphasize the importance of defining
the regions with relative accuracy as RAPT2 is less efficient than AM.
%

In Figure~\ref{fig:Dn_Banana_5} we plot the $D_n$ index averaged over the
$500$ chains replicates. One can  see that RAPT output, on average,
approximates the target CDF better than the other 3 algorithms, with RAPTOR
following very closely. The conclusions are similar to those based on MSE as 
AM and RAPT2 provide less accurate approximations than RAPT and RAPTOR.

In all the 4 algorithms, the proposal distribution uses the estimates
of the mixture's components variances.  In
Figure~\ref{fig:bananaMarginalVariancesEstimates} we show the average trend
of these estimates for the first coordinate of the state space. Here we see that
RAPTOR, RAPT and AM rapidly converge towards similar values, while RAPT2 gets
stuck on slightly smaller values.

In almost all diagnostics (acceptance rates, MSE, $D_n$ index), RAPTOR showed
performances very similar to those of RAPT. Indeed, the estimates of the region
specific means and variances resulted to be very similar in the two
algorithms. In Table~\ref{tab:bananaLocalEstimates} we report the average
final estimates of the mean and variances of the first coordinate in the two
regions, as well as the average estimated correlation between the first and the
second coordinate.
The value of the estimates shown in
Table~\ref{tab:bananaLocalEstimates}  help us determine that the 
partition  selected by RAPTOR is the same optimal partition we
have explicitely chosen for RAPT.
%

\section{Real Data Example: Genetic Instability of Eso\-pha\-geal Cancers}
\label{sec:real-data-exampleCancers}
We analyzed the ``Loss of Heterozygosity'' (LOH) dataset from the Seattle
Barrett's Esophagus research project \citep{barret06}, already analyzed in
\cite{warnes} and \cite{craiu-jeff-yang}, We refer to these papers and
references therein for a detailed description of the data. The dataset is composed by $40$ measures of frequencies of the event of
interest (LOH) with their associated sample sizes.  The model adopted for those
frequencies is a mixture model, as indicated by \cite{desai}:
\begin{equation}
  \label{eq:LOHModelObs}
  X_i \sim \eta\text{Binomial}(N_i, \pi_1) + (1-\eta)\text{Beta-Binomial}(N_i,
  \pi_2, \gamma)
\end{equation}

with priors:
\begin{equation}
  \label{eq:LOHModelPriors}
  \begin{aligned}
    \eta &\sim \text{Unif}[0,1],\\
    \pi_1 &\sim \text{Unif}[0,1],\\
    \pi_2 &\sim \text{Unif}[0,1],\\
    \gamma &\sim \text{Unif}[-30,30],\\
  \end{aligned}
\end{equation}
where $\eta$ is the probability of a location being a member of the binomial
group, $\pi_1$ is the probability of LOH in the binomial group, $\pi_2$ is the
probability of LOH in the beta-binomial group, and $\gamma$ controls the
variability of the beta-binomial group.  The parametrization adopted for the
Beta-Binomial distribution is such that $\gamma$'s range is the real line.
 As $\gamma \rightarrow -\infty$ the
beta-binomial becomes a binomial and as $\gamma \rightarrow \infty$ the
beta-binomial becomes a uniform distribution on $[0,1]$.  In order to facilitate the use of  the
RWM we have used the logistic transformation on the parameters
$\eta, \pi_1, \pi_2$.

We run 10 parallel chains of
the RAPTOR algorithm, allowing exchange of information between chains
 using INCA. The starting points for these
chains were drawn from a quasi-random distribution uniformly covering the
hypercube $[0.1, 0.9]^3 \times [-20, 20]$. All the chains were run for 200000
iterations, using the first 10000 as burn-in. The factor $\alpha$ which controls
the relative importance of the global vs. the local proposal jumps
 has been set to $0.7$.  In our experiments,  the RAPTOR chains displayed good performances even for smaller burn-in lengths
and different values of $\alpha$. However, setting a relatively big value of the
burn-in guarantees a less erratic behaviour of the chain between simulation
replications, while a relatively big value of $\alpha$ ensures a faster learning
of the relative importance of the two target mixture components.

\begin{figure}
  \centering
  \includegraphics[width=\textwidth]{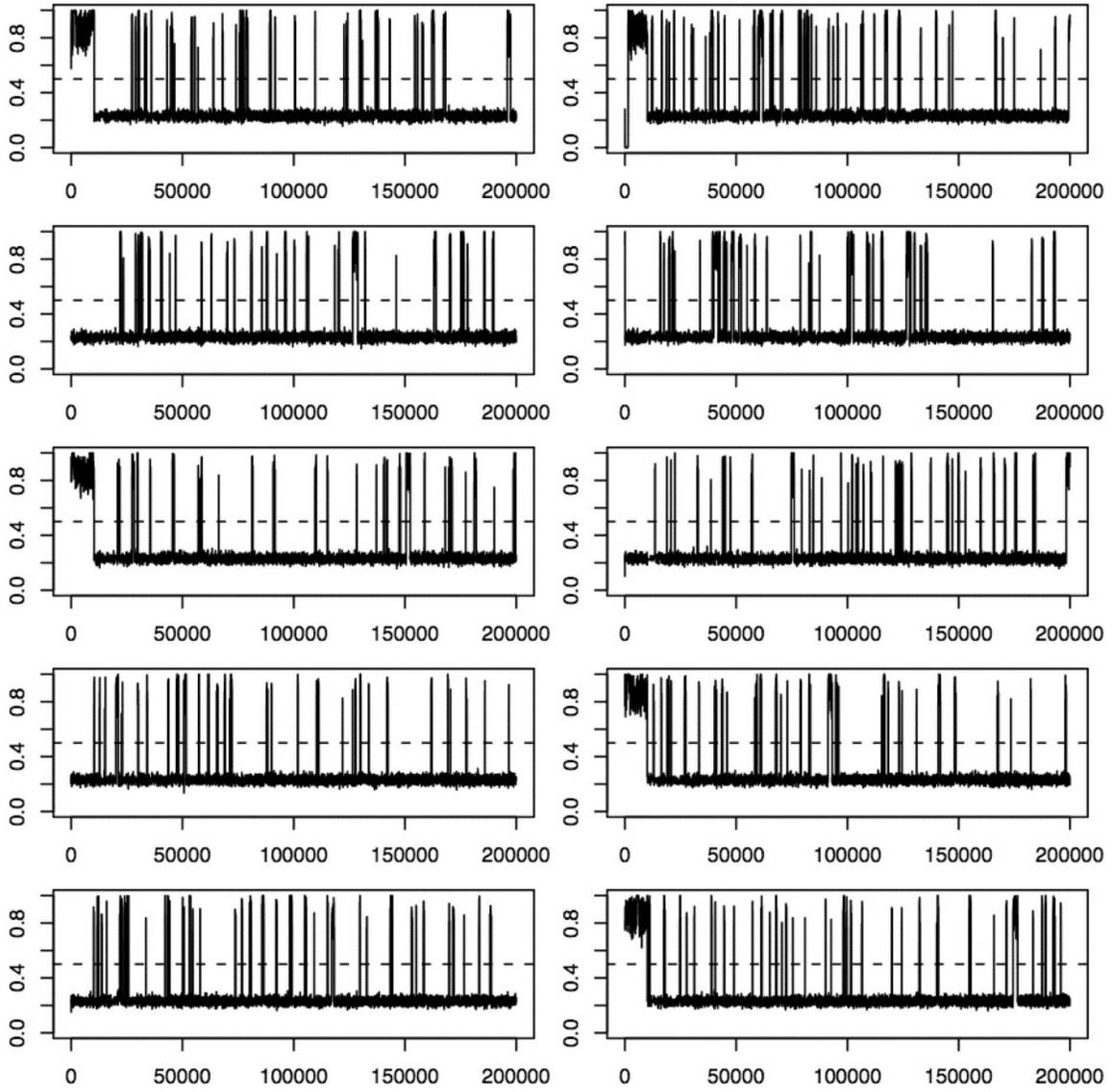}
  \caption{\em LOH data: parallel traces of $\pi_1$. Dotted horizontal line separates the two modes.}
  \label{fig:LOHp1Traces}
\end{figure}

In Figure~\ref{fig:LOHp1Traces} the traces of the coordinate $\pi_1$ of the 10
parallel chains are reported. Here one can see that all the chains switch very
often back and forth between the two posterior modes.

\begin{figure}
  \centering
  \includegraphics[width=0.7\textwidth]{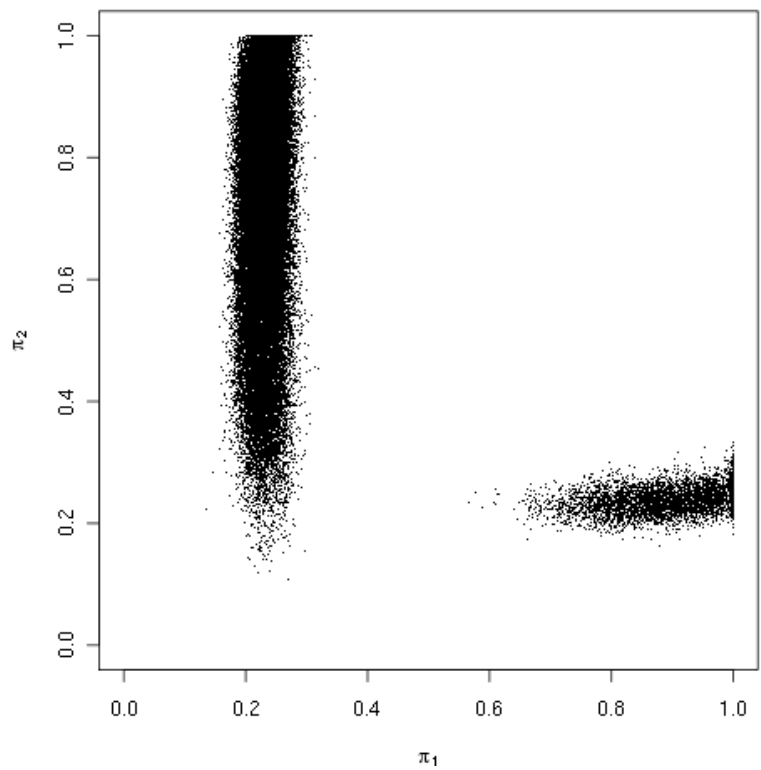}
  \caption{\em LOH data simulation: marginal scatterplot of $(\pi_1, \pi_2)$.}
  \label{fig:LOHScatterplot}
\end{figure}

In Figure~\ref{fig:LOHScatterplot} we show the marginal scatterplot of $(\pi_1, \pi_2)$
for all the samples obtained using the  $10$ parallel chains.
In this plot  the  differences between the mixture components of
the target distribution are clear.
In a situation like this, one single
setup for a RWM proposal distribution over the whole state
space would be highly inefficient, while a regional Adaptive Metropolis would
use different parameters values in each of the two regions.  Moreover, by using
RAPTOR, the regions can be identified automatically, without additional
input.  In the following, we
will label as region 1 the region with lower $\pi_1$ mean value, and as region 2
the region with bigger $\pi_1$ mean value.

\begin{figure}
  \centering\subfigure[]{
    \includegraphics[width=0.3\textwidth]{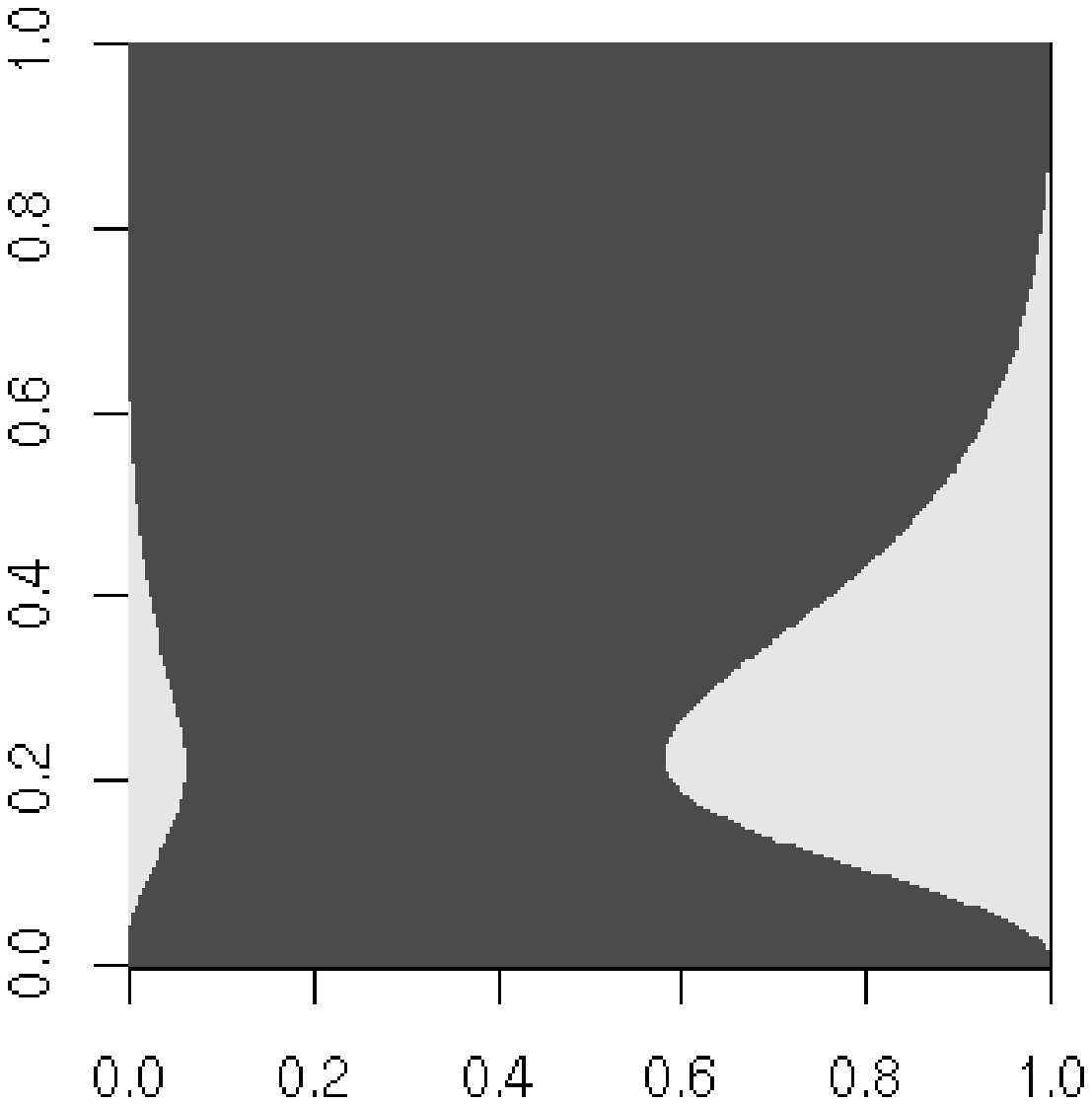}
    \label{fig:LOHBoundariesA}}
  \subfigure[]{
    \includegraphics[width=0.3\textwidth]{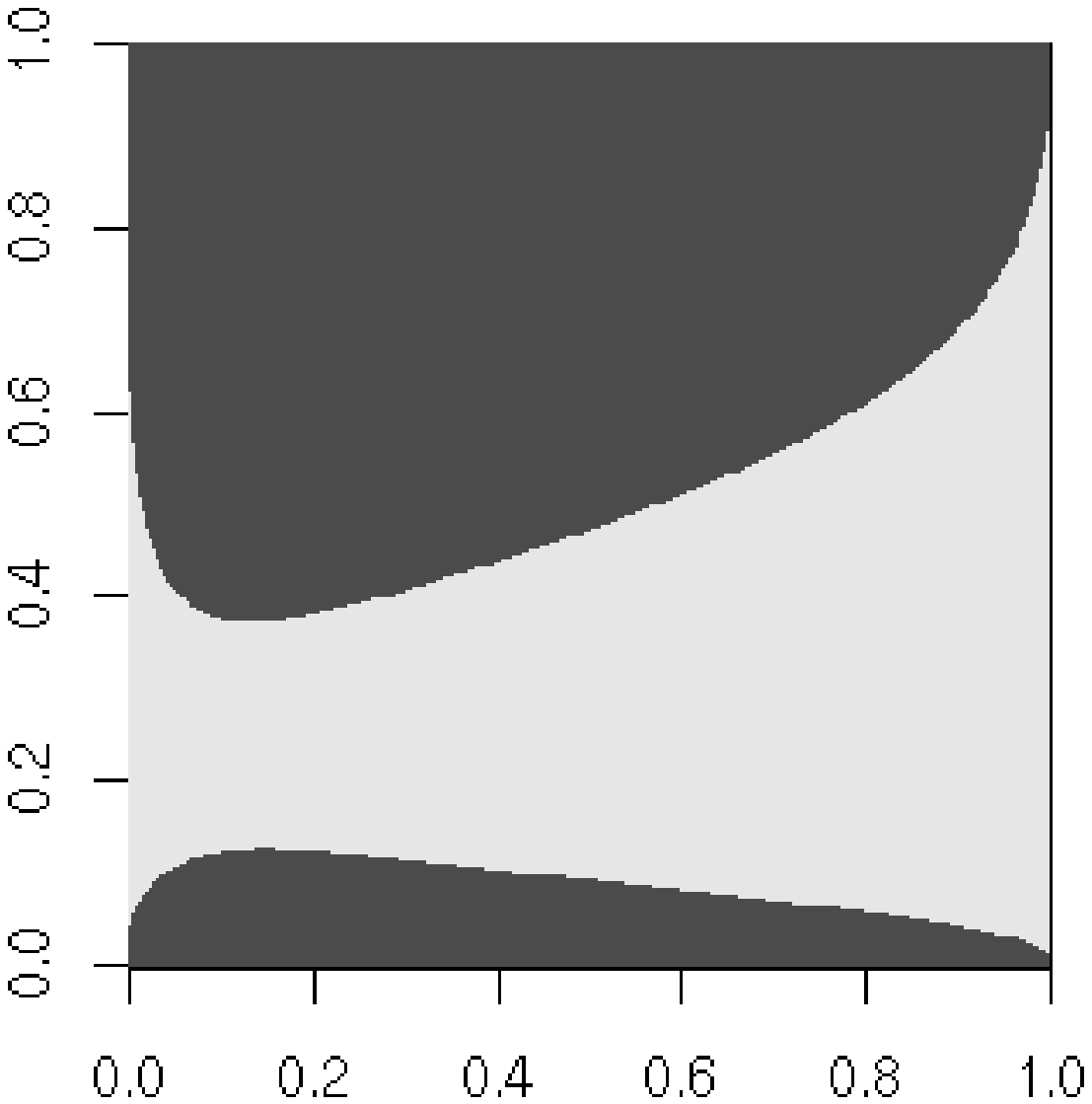}
    \label{fig:LOHBoundariesB}}
  \subfigure[]{
    \includegraphics[width=0.3\textwidth]{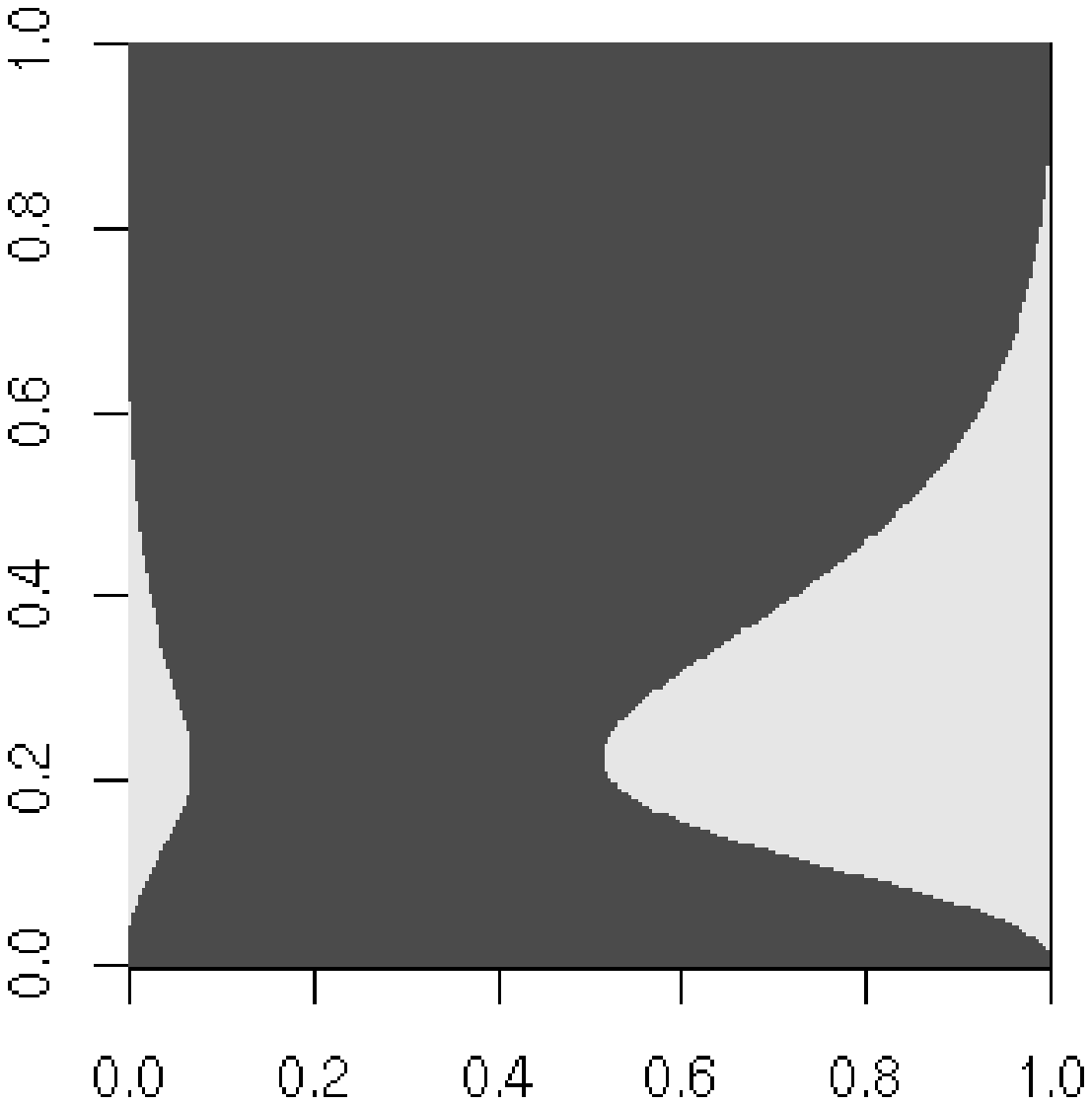}
    \label{fig:LOHBoundariesW}}
  \caption{\em LOH data simulation: slices of final RAPTOR boundaries estimates for
    values of $\eta$ and $\gamma$ equal to their mean in region 1 (left), region
    2 (center), whole state space (right).  Horizontal axis: $\pi_1$; vertical
    axis: $\pi_2$.  Dark gray: region 1; light gray: region 2.}
  \label{fig:LOHBoundaries}
\end{figure}

It is difficult to visualize the partition produced by RAPTOR in the four-dimensional space 
so instead we choose to show slices of the partition. In general, if the partition is defined according to (\ref{eq:boundary1})
 then for a fixed subset $I$ of the coordinates of interest and after fixing $x_I=(x_j: j \in I)$ at say, $\tilde{x}_I$ we can consider the slice through $\cas^k$ determined by  $\tilde{x}_I$ as
\begin{equation}
  \label{eq:boundary-slice}
  \begin{aligned}
    \cas^k(\tilde{x}_I) = \{x_{I^c}: \; \text{arg max}_{k'} N(x=(\tilde{x}_I,x_{I^c}); \mu_\eta^{k'},
    \Sigma_\eta^{k'}) = k\},
  \end{aligned}
\end{equation}
where $I^c$ is the complement of set $I$.
We can also define $\cas_{I^c}^k$ the projection of $\cas^k$ on the $x_{I^c}$-coordinate space and then 
$$\cas_{I^c}^k=\bigcup_{\tilde{x}_I} \cas^k(\tilde{x}_I),$$
where the union is taken over all the possible values of $\tilde{x}_I$.   One must choose which slices are more informative 
to look at and in general we choose $\tilde{x}$ to correspond to the local modes of $\pi$. In Figure~\ref{fig:LOHBoundaries} bi-dimensional slices of the RAPTOR regions
are plotted, for values for $\eta$ and $\gamma$ equal to their means in region 1
(Figure~\ref{fig:LOHBoundariesA}), region 2 (Figure~\ref{fig:LOHBoundariesB})
and in the whole state space (Figure~\ref{fig:LOHBoundariesW}). We can see that
region 2 is generally smaller than region 1, and that it gets a bigger area for
values of $\eta$ and $\gamma$ around their mean in that same region. In general,
it divides well the two posterios probability masses, allowing for an effective
application of the Regional Adaptive Metropolis scheme as implemented in RAPTOR.

In table~\ref{tab:LOHSummaries} we summarize final RAPTOR  estimates
on  the original  scales, and compare them  with the results reported
in~\cite{craiu-jeff-yang}. In this table we can see that the results are quite
similar, despite the different definitions of the boundary between
the two regions. This is probably due to the
fact that the two posterior modes are separated by a relatively large
region of low probability, so that a certain degree of variability in the boundary
specification is allowed, without affecting the results too much. However, it
must be noted here again that RAPTOR carries the  advantage that the boundary
has been learned \emph{automatically}, with no prior information input.
\begin{table}
  \centering
  \begin{tabular}{lrrr}
    \hline \hline
    & $S^1$ & $S^2$ & whole space \\
    \hline
    $\eta$ & 0.917 & 0.042 & 0.840 \\
    $\pi_1$ &  0.227 & 0.949 & 0.276 \\
    $\pi_2$ &  0.768 & 0.238 & 0.690 \\
    $\gamma$ & 12.187 & -13.249 & 10.336 \\
    \hline
  \end{tabular}\hspace{5mm}%
  \begin{tabular}{lrrr}
    \hline \hline
    & $S^1$ & $S^2$ & whole space \\
    \hline
    $\eta$ & 0.897 & 0.079 & 0.838 \\
    $\pi_1$ &  0.229 & 0.863 & 0.275 \\
    $\pi_2$ &  0.714 & 0.237 & 0.679 \\
    $\gamma$ & 15.661 & -14.796 & 13.435 \\
    \hline
  \end{tabular}
  \caption{\em Simulation results for LOH data. Region specific and global
    parameters means for RAPTOR (left) and RAPT (right).}
  \label{tab:LOHSummaries}
\end{table}

\section{Conclusions}

We propose a mixture-based approach for regional adaptation of the random walk Metropolis algorithm.  Using the theoretical foundations laid by \cite{Andrieu:2006rt} we use the online EM algorithm to adapt the parameters of a Gaussian mixtures using the stream of data produced by the MCMC algorithms. In turn, the mixture approximation is used within the regional adaptation paradigm defined by \cite{craiu-jeff-yang}. The main purpose of the current work is to provide a general method for defining a relatively accurate  partition of the sample space. Our simulations suggest that the approach produces partitions that are very close to the optimal one.

\renewcommand{\theequation}{\Alph{section}-\arabic{equation}}
\setcounter{equation}{0}  
\renewcommand{\thesection}{\Alph{section}}
\setcounter{section}{0}  

\section{The Online EM algorithm}
\label{sec:appendix:-line-em}

In the following, we will show how the Online EM algorithm presented in
\cite{Andrieu:2006rt} applies to the RAPTOR implementation presented in
section~\ref{sec:rec-strat-mixt-par-est}.

Consider the following exponential family mixture model
(cfr. \cite{Andrieu:2006rt},  pag. 1488):
\begin{equation}
  \label{eq:f_gamma}
  f_\eta(x, z) = \exp\{-\psi(\eta) + \langle T(x,z), \phi(\eta) \rangle\}\quad (\eta,
  x, z) \in \Omega \times \Xset \times \Zset
\end{equation}
where the density $f_\eta$ is defined w.r.t. some convenient measure on
$\Omega \times \Xset \times \Zset$.  $T(x, z)$ is the complete data sufficient
statistic, i.e. the likelihood $f_\eta$ is a function of the complete data
pair $(x,z)$ only through the statistic $T(x, z)$. Denote with $\tilde
q_\eta(x)$ the marginal density of $f_\eta$:
\begin{equation}
  \label{eq:tilde_q_gamma}
  \tilde q_\eta(x) = \int_\Zset f_\eta(x, z) \mu(dz)
\end{equation}
One special case of~\eqref{eq:f_gamma} is the finite mixture of Gaussians:
\begin{equation}
  \label{eq:gauss_mix_appendix}
  f_\eta(x, z) = \prod_{k=1}^K \left [ \beta_\eta^k N(x; \mu_\eta^k, \Sigma_\eta^k) \right ]^{\mathbf{1}(z=k)}
\end{equation}

It can be verified that~\eqref{eq:gauss_mix_appendix} is indeed a special case
of~\eqref{eq:f_gamma} where $T(x, z)$ takes the form:
\begin{equation}
  \label{eq:Txz_gaussmix}
  T(x,z) = \{ \indicator\{z=k\} \cdot (1,\; x,\; xx^T), \quad k=1,\ldots,K \}
\end{equation}
In the above, $x$ denotes a column vector of dimension $d$ and $x x^T$ is
the usual matrix product.  The marginal density $\tilde q_\eta(x)$ takes the
well known form:
\begin{equation}
  \label{eq:tilde_q_gamma_gaussmix}
  \tilde q_\eta(x) = \sum_{k=1}^K \beta_\eta^k N(x; \mu_\eta^k, \Sigma_\eta^k)
\end{equation}

The central point in the classical EM algorithm is estimating the expected value
of the complete log-likelihood \eqref{eq:f_gamma} conditional on $X$ and a value
$\eta'$ of $\eta$.  Thus we need to estimate
\begin{equation}
  \label{eq:E_log_f_gamma_X_Z}
  \mathbb{E}\{\text{log}(f_\eta(X, Z)) |X, \eta' \}
\end{equation}
Since the complete log-likelihood is linear in $T(X,Z)$ this reduces to
computing the conditional expected value of the sufficient statistic.
We start by deriving the conditional distribution of $Z$ given $X$ and $\eta$:
\begin{equation}
  \label{eq:nu_gamma_xz}
  \begin{aligned}
    \nu_\eta(x, z) & := \frac{f_\eta(x, z)}{\tilde q_\eta(x)}\\
     & = \frac{\beta_\eta^z N(x; \mu_\eta^z, \Sigma_\eta^z)}{
       \sum_{k=1}^K \beta_\eta^k N(x; \mu_\eta^k, \Sigma_\eta^k)
     }
  \end{aligned}
\end{equation}

Now we can define:
\begin{equation}
  \label{eq:nu_gamma_T_x}
  \begin{aligned}
    \nu_\eta T(x) & := \int_\Zset T(x,z) \nu_\eta(x, z) \mu(dz) \\
    & = \sum_{k=1}^K \nu_\eta(x, k)\; T(x, k)
  \end{aligned}
\end{equation}
which is the expected value of the complete data sufficient statistic given $X$
and $\eta$.  This can be estimated recursively by the following stochastic
approximation scheme:
\begin{equation}
  \label{eq:stoch_approx_thetak}
  \theta_{n+1}  = (1 - \alpha_{n+1}) \theta_n + \alpha_{n+1} \nu_{\eta_n}T(X_{n+1})
\end{equation}
where $\theta_n \in \Theta = T(\Xset, \Zset)$.  The Maximization Step is the
same as that in the classical EM setup\begin{equation}
  \label{eq:EM_maximization}
  \begin{aligned}
    \beta_{\eta_n}^k & = \theta_n^{0,k} \\
    \mu_{\eta_n}^k & = \frac{\theta_n^{1,k}}{\theta_n^{0,k}} \\
    \Sigma_{\eta_n}^k & = \frac{\theta_n^{2,k}}{\theta_n^{0,k}} - \mu_{\eta_n}^k
    \mu_{\eta_n}^{k\prime}
  \end{aligned}
\end{equation}
where we have expressed $\theta$ as:
\begin{equation}
  \label{eq:appendix_theta_explosion}
  \theta = \{(\theta^{0,k},\; \theta^{1,k},\; \theta^{2,k}), \quad k = 1, \ldots, K\}
\end{equation}
with $\theta^{0,k} \in \Rset$, $\theta^{1,k} \in \Rset^d$, $\theta^{2,k}
\in \Rset^{d \times d}$.  Different choices are possible for the
learning weights $\alpha_n$. One possibility which guarantees convergence is
$\alpha_n = n^{-1}$, and this is indeed what is used in the RAPTOR
implementation.

\bibliography{superref}
\bibliographystyle{ims}

\end{document}